\documentclass[twocolumn,floatfix,amssymb,aps,prb,nobibnotes,nofootinbib,superscriptaddress]{revtex4-2}
\usepackage{docs}
\usepackage{tikz}
\usepackage{booktabs}
\usepackage{graphicx}
\usepackage{float}
\usepackage{amsmath}
\usepackage{pifont}

\usepackage[colorlinks=true, linkcolor=blue, citecolor=red, urlcolor=magenta]{hyperref}

\expandafter\ifx\csname package@font\endcsname\relax\else
 \expandafter\expandafter
 \expandafter\usepackage
 \expandafter\expandafter
 \expandafter{\csname package@font\endcsname}%
\fi
\DeclareRobustCommand\substyle{\name@idx{document substyle}}%
\DeclareRobustCommand\classoption{\name@idx{document class option}}%
\DeclareRobustCommand\classname{\name@idx{document class}}%
\def\name@idx#1#2{%
 {\ttfamily#2}%
 \index{#2\space#1=\string\ttt{#2}\space#1}\index{#1>#2=\string\ttt{#2}}%
}%

\begin{document}
\title{Quantum Mpemba Effects from Symmetry Perspectives}
	
\author{Hui Yu}
\affiliation{Beijing National Laboratory for Condensed Matter Physics and Institute of Physics, Chinese Academy of Sciences, Beijing 100910, China}

\author{Shuo Liu}
\affiliation{Institute for Advanced Study, Tsinghua University, Beijing 100084, China}

\author{Shi-Xin Zhang}
\email{znfesnpbh@gmail.com}
\affiliation{Beijing National Laboratory for Condensed Matter Physics and Institute of Physics, Chinese Academy of Sciences, Beijing 100910, China}

\date{\today}%


\begin{abstract}
Non-equilibrium dynamics have become a central research focus, exemplified by the counterintuitive Mpemba effect where initially hotter systems can cool faster than colder ones. Studied extensively in both classical and quantum regimes, this phenomenon reveals diverse and complex behaviors across different systems. This review provides a concise overview of the quantum Mpemba effect (QME), specifically emphasizing its connection to symmetry breaking and restoration in closed quantum many-body systems. We begin by outlining the classical Mpemba effect and its quantum counterparts, summarizing key findings. Subsequently, we introduce entanglement asymmetry and charge variance as key metrics for probing the QME from symmetry perspectives. Leveraging these tools, we analyze the early- and late-time dynamics of these quantities under Hamiltonian evolution and random unitary circuits. We conclude by discussing significant challenges and promising avenues for future research. 
\end{abstract}
\maketitle

\section{Introduction}
The Mpemba effect is a fascinating and counterintuitive phenomenon in which a hotter system can freeze faster than a cooler one under identical conditions. The phenomenon was notably brought to modern scientific attention in the 1960s by a student who observed that hot ice cream mixture froze quicker than a colder one \cite{mpemba1969cool}. The effect defies the conventional thermodynamics expectations, which suggest that a hotter system should always take longer to cool to a given temperature than a colder one. Such effects have been debated for centuries, with historical accounts \cite{bacon1962opus,descartes1948discours,groves2009now} dating back to ancient times, and it continues to intrigue scientists due to its complex and elusive nature \cite{burridge2016questioning,burridge2020observing}. 

In classical systems, the Mpemba effect has been observed in a variety of contexts, each offering unique insights into its underlying mechanisms. One prominent example is supercooling \cite{auerbach1995supercooling,esposito2008mpemba}, where a substance is cooled below its freezing point without undergoing solidification. In such cases, the system may remain trapped in a metastable phase for an extended period before abruptly transitioning to the equilibrium state. Beyond supercooling, the effect has also been studied in granular fluids \cite{lasanta2017hotter}, where the interplay between particle interactions and energy dissipation can lead to anomalous cooling behaviors.
Similarly, in clathrate hydrate \cite{ahn2016experimental} and carbon nanotube resonators \cite{greaney2011mpemba}, unexpected cooling patterns have been reported.  Moreover, the Mpemba effect has been experimentally observed in classical Markovian systems, where stochastic system-bath interactions drive the anomalous relaxation dynamics \cite{kumar2020exponentially,bechhoefer2021fresh}. Theoretical studies have identified distinct mechanisms behind this phenomenon: Ref.~\cite{lu2017nonequilibrium} shows that the effect arises when a hotter initial state has substantially smaller overlap with the system’s slowest-decaying eigenmodes compared to a colder state, enabling faster equilibration. In Ref.~\cite{klich2019mpemba}, the authors introduce the strong Mpemba effect, where relaxation is exponentially accelerated for specific initial temperatures. This effect is characterized by the Mpemba index,  an integer that counts the number of such special initial temperatures. Notably, the parity of the index (odd or even) is a topological invariant of the system, ensuring robustness against small parameter variations. They demonstrate that this effect persists even in the thermodynamic limit and is closely linked to thermal overshooting—a phenomenon in which the temperature of the relaxing system exhibits non-monotonic decay over time. Despite these diverse observations, a universally accepted explanation for the classical Mpemba effect is still absent, with proposed mechanisms ranging from differences in thermal conductivity and evaporation rates to more intricate considerations of non-equilibrium dynamics. This phenomenon continues to be an active area of research within the scientific community \cite{hu2018conformation,keller2018quenches,holtzman2022landau,baity2019mpemba,kumar2022anomalous,santos2024mpemba,schwarzendahl2022anomalous,teza2023relaxation,pemartin2024shortcuts,gal2020precooling,walker2022mpemba,walker2023optimal,bera2023effect,teza2025speedups}.

In recent years, the focus of the Mpemba effect has expanded beyond classical physics, extending into the intriguing realm of quantum systems, which are governed by the principles of quantum mechanics. In quantum regimes, the Mpemba effect can manifest in two distinct ways: through open quantum systems that interact with an external environment or through isolated quantum systems that evolve under unitary dynamics. Ref.~\cite{nava2019lindblad} explores the Mpemba effect in open quantum systems using a quantum Ising model coupled to a thermal bath, revealing that metastable phases enable Mpemba-like behavior. Despite initial temperature differences, the hotter system reached equilibrium faster after a thermal quench, showing transient magnetization dynamics governed by Lindblad evolution. This confirms that quantum effects preserve rather than disrupt the Mpemba phenomenon in supercooled systems. Given the prevalence of the Mpemba effect in classical Markovian systems, a natural question arises: Can this phenomenon also occur in open quantum Markovian systems? In Markovian systems described by Lindblad dynamics, researchers discovered that a unitary rotation can suppress the slowest-decaying eigenmode, leading to exponentially accelerated relaxation—a signature of the strong Mpemba effect 
\cite{carollo2021exponentially}. Ref.~\cite{zhang2025observation} presents a protocol for generating the strong Mpemba effect by implementing precisely controlled rotational operations on a single trapped ion initialized in its ground state. However, faster relaxation alone doesn't guarantee the Mpemba effect; the system must also start sufficiently far from equilibrium, as shown through free-energy analysis of specially prepared initial states \cite{moroder2024thermodynamics}. Variants like the inverse Mpemba effect (where a colder system heats up faster) using a single trapped ion qubit, provide experimental evidence of this phenomenon in a quantum system \cite{aharony2024inverse}, while mixed versions of Mpemba effect, where relaxation speed depends on both thermal and non-thermal effects, appear in quantum dots due to intermediate relaxation modes \cite{chatterjee2023quantum}. Non-Hermitian systems exhibit even richer relaxation dynamics, with multiple observable crossings—such as ground-state energy, entropy, and distance to the steady state—emerging during thermalization \cite{chatterjee2024multiple}. Recent studies \cite{kochsiek2022accelerating,ivander2023hyperacceleration,wang2024mpemba,longhi2024photonic,longhi2024bosonic,liu2024speeding,boubakour2025dynamical,furtado2024strong,qian2024intrinsic,dong2025quantum,wang2024mpemba2,kheirandish2024mpemba} have extensively investigated the Mpemba effect in other open Markovian quantum systems, analyzing its emergence under different conditions. Beyond Markovian regimes, non-Markovian dynamics introduce new possibilities like the extreme quantum Mpemba effect \cite{strachan2024non}, where systems exploit memory effects to achieve ultra-fast equilibration. Ref.~\cite{wang2024going} develops a non-Markovian exact master equation to study quantum system dynamics, revealing how memory effects and system-bath correlations lead to deviations from standard Markovian relaxation. 

\begin{figure*}[htb]
\includegraphics[scale=0.75]{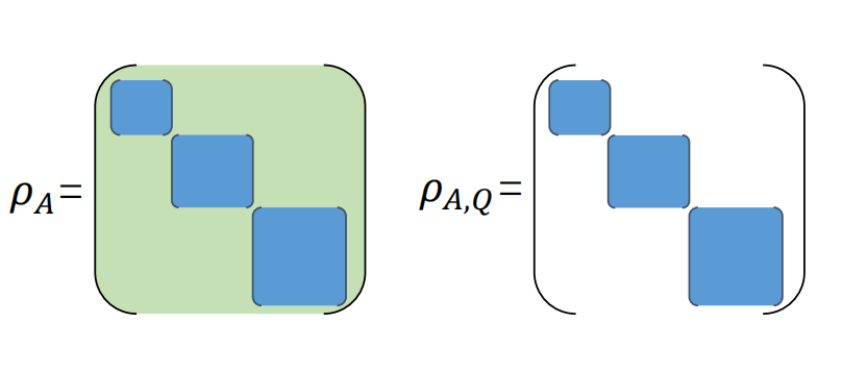}
\caption{The reduced density matrix of subsystem $A$, denoted as $\rho_{A}$, and its symmetry-projected counterpart, $\rho_{A,Q}$, are illustrated in the definition of entanglement asymmetry. Each blue block corresponds to a specific $U(1)$ symmetry sector with a fixed charge $Q$. In $\rho_{A}$, the light green background highlights off-diagonal matrix elements between different charge sectors. In contrast, $\rho_{A,Q}$ (white background) contains no off-diagonal elements between different $Q$ sectors.}
\label{fig:density matrix}
\end{figure*}

The quantum Mpemba effect (QME) \cite{ares2025quantum} reveals a remarkable non-equilibrium phenomenon in isolated quantum systems: symmetry restoration occurs faster in states that initially exhibit stronger symmetry breaking. This closed-system manifestation of QME is especially intriguing because relaxation dynamics are governed by intrinsic quantum fluctuations, unlike in open quantum systems, where environmental dissipation typically dominates. This effect emerges following a global quantum quench, where the system is prepared in a pure, non-equilibrium state and evolves under unitary dynamics. Recent studies have reported the QME in both integrable and chaotic quantum systems \cite{ares2023entanglement,liu2024symmetry,turkeshi2024quantum}, with notable examples including the rapid restoration of U(1) symmetry when quenching from highly asymmetric initial states under a U(1)-symmetric Hamiltonian \cite{fagotti2014relaxation,essler2016quench,doyon2020lecture,fagotti2014conservation,bertini2015pre,vidmar2016generalized,alba2021generalized,calabrese2016quantum,polkovnikov83nonequilibrium,bastianello2022introduction,ares2023lack}. These findings have since been extended to various other settings. Notably, Refs.~\cite{khor2024confinement,rylands2024dynamical,murciano2024entanglement,ferro2024non} explore dynamical symmetry restoration and non-equilibrium quantum dynamics in one-dimensional spin chains, with particular focus on phenomena like the QME and the role of integrability. The QME has also been discovered in strongly disordered, many-body localized systems (MBL) \cite{liu2024quantum}. This study extends the understanding of anomalous thermalization dynamics to disordered, non-ergodic systems, revealing how MBL alters the interplay between symmetry and relaxation. The imaginary-time QME \cite{chang2024imaginary}, where 
certain initial states converge faster to the ground state under imaginary-time evolution compared to others, has been reported in various physical systems. Ref.~\cite{guo2025skin} reveals that the QME can even be realized in non-Hermitian systems. Recent studies \cite{ares2025entanglement,klobas2024translation} have extended investigations of anomalous relaxation to random quantum circuits, revealing universal dynamical features like symmetry restoration and characteristic relaxation scaling. Furthermore, Ref.~\cite{di2025measurement} identifies a measurement-induced QME in monitored quantum systems. Additionally, Refs.~\cite{yu2025symmetry,yu2025tuning} systematically examine the QME under non-integrable Hamiltonian evolution.  Extensions to higher-dimensional systems have also revealed new insights into entanglement propagation and thermalization \cite{yamashika2024entanglement,yamashika2025quenching}. A microscopic mechanism for the QME in generic integrable systems has been proposed, linking it to quasiparticle excitations \cite{rylands2024microscopic}. Furthermore, the QME has been experimentally demonstrated in trapped-ion quantum simulators \cite{joshi2024observing}. Remarkably, this effect persists even in the presence of experimental imperfections like noise and disorder. This experimental realization provides crucial insights: when symmetry restoration is driven purely by environmental decoherence, no QME crossings occur, establishing that intrinsic quantum fluctuations are essential for the effect. In this review, we examine the QME through the lens of symmetry breaking and restoration, highlighting its manifestation in both Hamiltonian and random circuit dynamics. Our discussion underscores the QME’s significance in advancing the understanding of non-equilibrium quantum dynamics. 

The remainder of this review is structured as follows. In Section II, we introduce two complementary diagnostic measures - entanglement asymmetry and charge variance - to systematically quantify how far a system deviates from equilibrium. Section III investigates the emergence of the QME in scenarios where subsystem symmetry is restored at late times. In Section IV, we explore the robustness of the QME when subsystem's symmetry remains broken at late times. Finally, Section V concludes by outlining key unsolved questions and promising future directions in the study of non-equilibrium quantum phenomena. 

\section{The metrics}
Many potential candidates exist to quantify the distance of a system to its equilibrium state, but three have been particularly prominent in the study of quantum dynamics. The quantum relative entropy, defined as $S(\rho(t)||\rho_{eq})=\text{Tr}[\rho(t)(\log\rho(t)-\log\rho_{eq})]$, compares the time-evolved density matrix $\rho(t)$ with the steady-state equilibrium $\rho_{eq}$ and is closely tied to non-equilibrium free energy. It has been utilized in investigations of the QME \cite{moroder2024thermodynamics,chatterjee2023quantum}. Another key measure, the trace distance, defined as $d_{\text{Tr}}(\rho(t))=\frac{1}{2}\text{Tr}\sqrt{A^{\dagger}A}$, where $A=\rho(t)-\rho_{eq}$, is particularly useful due to its monotonic decay under Markovian dynamics. This property has enabled experimental studies of strong and inverse QMEs \cite{zhang2025observation,aharony2024inverse}. In contrast, the Frobenius distance, defined as $d_{F}(\rho(t))=\sqrt{\text{Tr}A^{\dagger}A}$, offers a simpler computational alternative to the trace distance, though it lacks monotonicity in Markovian systems \cite{carollo2021exponentially}. 

While these measures offer different insights into equilibration, recent work suggests that entanglement asymmetry is a practical, easy-to-measure proxy for the distance to an equilibrium state. This measure originates from the study of entanglement properties in quantum many-body systems and has been widely employed to quantify symmetry breaking in both quantum field theories \cite{capizzi2024universal,chen2024renyi,capizzi2023entanglement} and out-of-equilibrium many-body systems \cite{rylands2024microscopic,khor2024confinement,ares2025entanglement}. To define this quantity, we typically partition the entire system into a subsystem $A$ and its complement $\overline{A}$. The reduced density matrix for subsystem $A$, denoted as $\rho_{A}$, is obtained by tracing out the degrees of freedom in $\overline{A}$, expressed as $\rho_{A}=\text{Tr}_{\overline{A}}(\rho)$. Here, $\rho$ represents the density matrix for the entire system. The entanglement asymmetry (EA) \cite{ares2023entanglement} is defined as:
\begin{eqnarray}
    \Delta S_{A}^{n}  = S^{n}(\rho_{A, Q}) - S^{n}(\rho_{A}),
\end{eqnarray}
where $S^{n}(\rho)=\frac{1}{1-n}\log \text{Tr}(\rho^{n})$ is the $n$-th R\'{e}nyi entropy. In the limit $n\rightarrow 1$, the R\'{e}nyi entropy reduces to the Von Neumann entanglement entropy, given by $S(\rho)=-\text{Tr}(\rho\log\rho)$. Here, $\rho_{A, Q} = \sum_{q \in \mathbb{Z}} \Pi_{q} \rho_{A} \Pi_{q}$ where $\hat{Q}_{A} = \sum_{i \in A} \sigma_{i}^{z}$ in case of U(1) symmetry and $\Pi_{q}$ is the projector onto eigenspace of $\hat{Q}_{A}$ with charge $q$. Consequently, $\rho_{A, Q}$ is block diagonal in the eigenbasis of $\hat{Q}_{A}$. A schematic representation of $\rho_{A}$ and $\rho_{A,Q}$ is illustrated in Fig.~\ref{fig:density matrix}. The EA satisfies two key properties: (1) $\Delta S_{A} \geq 0$ since the EA is defined as the relative entropy between $\rho_{A, Q}$ and $\rho_{A}$. (2) $\Delta S_{A} = 0$ if and only if $\rho_{A, Q}=\rho_{A}$. Therefore, EA quantifies the degree of symmetry breaking in subsystem $A$ by measuring the distinguishability between the reduced density matrix $\rho_{A}$ and its symmetry-projected counterpart $\rho_{A,Q}$. Note that the symmetry for subsystem mixed states investigated here corresponds to the concept of weak symmetry in Refs.~\cite{lessa2025strong,sala2024spontaneous}.

In the experimental implementation, EA is measured by determining the  $S^n(\rho_{A,Q})$ of the subsystem state. The first method \cite{joshi2024observing} involved obtaining the subsystem density matrix $\rho_A$ via quantum state tomography, leveraging the classical shadow framework \cite{huang2020predicting}. The required block-diagonal state $\rho_{A, Q}$ was then computed numerically from $\rho_A$ via projection. An alternative protocol prepares the required block-diagonal state $\rho_{A,Q}$ by applying local random unitary gates $U$ that preserve the symmetry structure. This operation effectively dephases the subsystem by decohering the off-block-diagonal elements of the density matrix, directly yielding a state on quantum hardware in the desired block-diagonal form $\rho_{A,Q}$. The entropy $S^n(\rho_{A,Q})$ of this prepared state was then measured using randomized measurements \cite{brydges2019probing} on the subsystem.
Both protocols exhibit an inherent exponential complexity with respect to the subsystem size for general states.

In parallel with the EA analysis, we also compute the charge variance (CV), defined as 
\begin{eqnarray}
\sigma^{2}_{Q}= \langle \hat{Q}^{2} \rangle - \langle \hat{Q} \rangle^{2},
\end{eqnarray}
where $\hat{Q} = \sum_{i=1}^{L} \sigma_{i}^{z}$ represents the total charge operator for a system of size $L$ and the expectation value $\langle \hat{Q}\rangle$ is defined as $\langle \hat{Q}\rangle=\text{Tr}(\rho \hat{Q})$. If the expectation value $\langle \cdot \rangle$ is taken with respect to a U(1)-symmetric state where the state resides entirely within a single charge sector, then $\sigma^{2}_{Q}=0$. Conversely, if the state's charge distribution spans multiple sectors, the charge variance becomes nonzero. CV can be measured directly on the computational basis, which is simpler than EA measurement protocols discussed above.

The quantum Mpemba effect describes a scenario where a system initially ``further'' from a target state (e.g., hotter, or with more pronounced symmetry breaking) can reach it faster than a system that started ``closer''.
More formally, consider two identical quantum systems prepared in distinct initial states, described by density matrices $\rho_{1}(0)$ and $\rho_{2}(0)$. Let $O(t)$ be a relevant physical observable characterizing the system's proximity to equilibrium or a symmetric state. If, without loss of generality, $O(\rho_{1}(0)) > O(\rho_{2}(0))$, the QME is said to occur if there exists a characteristic time $t_{\text{QME}}$ such that this inequality inverts for subsequent times, i.e., $O(\rho_{1}(t)) < O(\rho_{2}(t))$ for $t > t_{\text{QME}}$.

This general definition applies to various metrics. For instance, when using entanglement asymmetry, if $\Delta S^{n}(\rho_{1}(0)) > \Delta S^{n}(\rho_{2}(0))$, namely state $\rho_{1}(0)$ exhibits stronger initial symmetry breaking than $\rho_{2}(0)$, the QME manifests when the system evolving from $\rho_{1}(0)$ achieves a lower EA value at later times, i.e., $\Delta S^{n}(\rho_{1}(t)) < \Delta S^{n}(\rho_{2}(t))$ for $t > t_{\text{QME}}$.
Similarly, for charge variance, if initially $\sigma^{2}_{Q}(\rho_{1}(0)) > \sigma^{2}_{Q}(\rho_{2}(0))$, the QME in CV occurs when $\sigma^{2}_{Q}(\rho_{1}(t)) < \sigma^{2}_{Q}(\rho_{2}(t))$ for $t > t_{\text{QME}}$.

The crossing timescale $t_{\text{QME}}$ depends on the initial states, the specific observable, and the underlying physical model. In many contexts, such as EA dynamics during symmetry restoration in chaotic systems, this crossing occurs at relatively early times, potentially on the order of the subsystem size.
Furthermore, by comparing the initial ordering of the observable with its ordering in the late-time steady state, one can infer the parity of the number of crossings: if the initial order $O(\rho_1(0)) > O(\rho_2(0))$ inverts to $O(\rho_1(t\to\infty)) < O(\rho_2(t\to\infty))$, then for continuous evolution, an odd number of crossings (at least one for QME) must occur. Conversely, if the initial order is preserved at late times, an even number (could be zero) of crossings is implied.

\section{Symmetry restoration dynamics}
\subsection{U(1)-asymmetric initial states with U(1)-symmetric random circuit}
\begin{figure}[htbp]
\begin{center}
\includegraphics[scale=0.19]{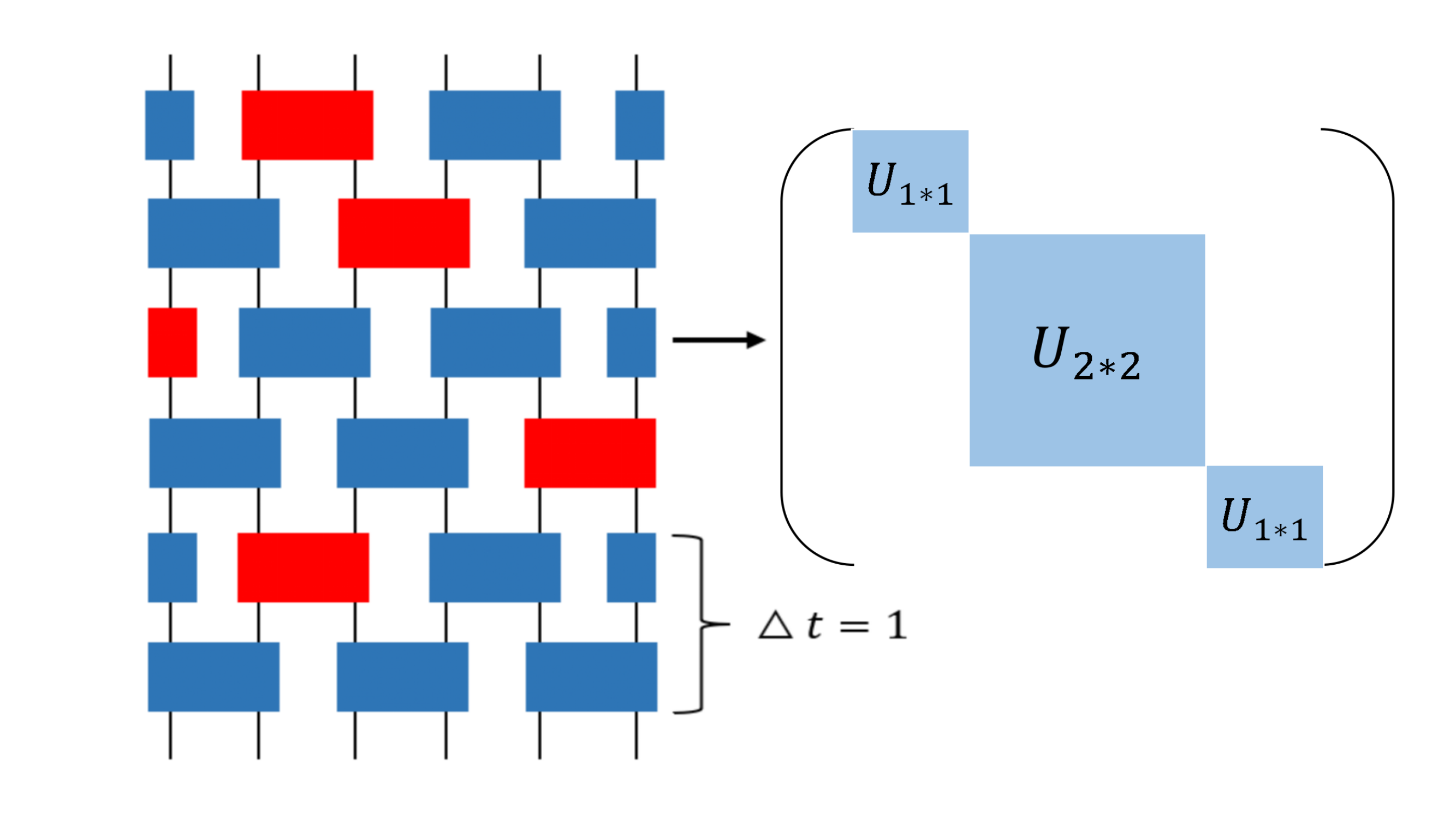}
\caption{Schematic illustration of a 6-qubit non-symmetric random circuit with periodic boundary conditions. Gates are arranged in the even-odd brick-wall pattern. The blue and red rectangles represent U(1)-symmetric and random Haar gates, respectively. The basis for the U(1)-symmetric gate is listed in the following order: $\vert 00 \rangle$, $\vert 01 \rangle$, $\vert 10 \rangle$ and $\vert 11 \rangle$.}
\label{fig:non-symm cir}
\end{center}
\end{figure}

The random circuit \cite{fisher2023random,li2019measurement,bao2020theory,zabalo2020critical,jian2020measurement}, depicted in Fig.~\ref{fig:non-symm cir}, is constructed from two-qubit random U(1)-symmetric gates and random Haar gates, arranged in a brick-wall pattern. The U(1)-symmetric gates are designed to preserve the charge sector, with their matrix form explicitly illustrated in Fig.~\ref{fig:non-symm cir}. Each block within these gates is randomly drawn from the Haar measure \cite{li2023d,hearth2023unitary,li2023designs}. Importantly, due to the U(1)-symmetric nature of these gates, there are no matrix elements connecting different charge sectors, which enforces the conservation of charge within each sector. The degree of symmetry breaking is governed by the doping probability $P_{\text{Haar}}$, which sets the probability of replacing a U(1)-symmetric gate with a random Haar gate at any given site. The fully symmetric case, $P_{\text{Haar}}=0$, corresponds to a U(1)-symmetric random circuit.

The time evolution of the circuit is discretized, with each time step corresponding to the application of two consecutive layers of gates. The unitary operator $U$ governing the evolution between two consecutive time steps $t$ and $t+1$ is defined as: 
\begin{eqnarray}
U = (U_{1,L}U_{L-1,L-2}...U_{2,3})(U_{L,L-1}U_{L-2,L-3}...U_{1,2}) \label{unitary_onestep}
\end{eqnarray}
where the second (first) bracket represents the operations applied in the first (second) layer of the time step. Here, $U_{i,i+1}$ denotes either a $U(1)$-symmetric gate or a random Haar gate acting between qubits $i$ and $i+1$.
The state $\vert \psi (t+1) \rangle$ at time $t+1$ is updated from the state $\vert \psi (t) \rangle$ through the action of the unitary operator $U$ as:  
\begin{eqnarray}
\vert \psi(t+1) \rangle = U \vert \psi (t) \rangle.\label{wf_update}
\end{eqnarray}
This process is iterated to evolve the state over multiple time steps. And $\mathbb{E} [\Delta S_{A}^{n}]$ is computed by averaging $\Delta S_{A}^{n}$ over different circuit configurations. In the following study, we consider three initial states: the ferromagnetic state $\vert 000...0 \rangle$, the antiferromagnetic state $\vert 0101..01 \rangle$ and the domain-wall state $\vert 000..111 \rangle$, where the domain wall is positioned at the center of the chain. To incorporate the effect of symmetry breaking in the initial state, we introduce tilted ferromagnetic states defined as:
\begin{eqnarray}
    \vert \psi_{i} (\theta)\rangle =  e^{-i\frac{\theta}{2} \sum_{j} \sigma_{j}^{y}} \vert 000...0 \rangle,
    \label{eq: tilted ferromagnetic}
\end{eqnarray}
where $\sigma_{j}^{y}$ is the Pauli-$y$ matrix acting on the $j$-th qubit, and $\theta$ is a tuning parameter that controls the strength of symmetry breaking in the initial state. When $\theta=0$, Eq.~\eqref{eq: tilted ferromagnetic} is U(1)-symmetric, resulting in a vanishing EA. As $\theta$ increases, the EA grows, reaching its maximum value at $\theta=\pi/2$. The tilted antiferromagnetic and tilted domain wall states are constructed in a similar manner. 

\begin{figure}[htbp]
\begin{center}
\includegraphics[width=0.47\textwidth, keepaspectratio]{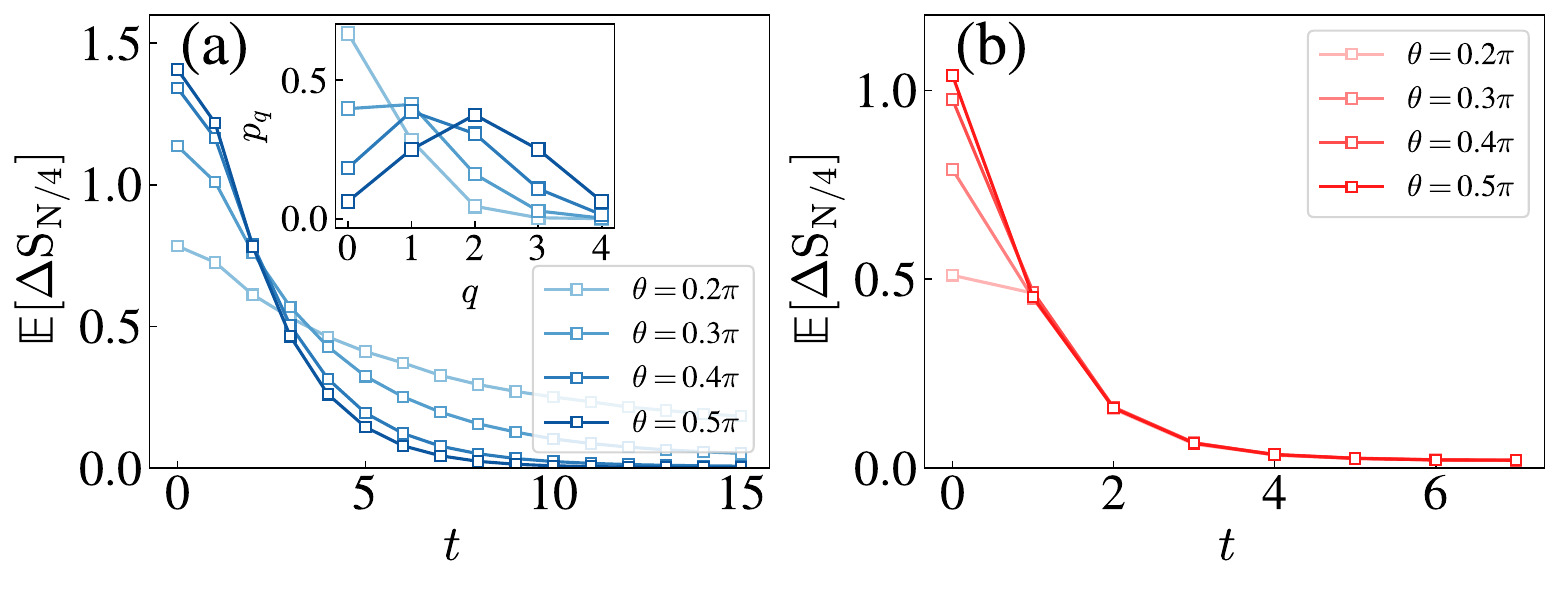}
\caption{Panels (a) and (b) show the EA dynamics of subsystem $A=[0, N/4]$ for random quantum circuits with $P_{\text{Haar}}=0$ and $P_{\text{Haar}}=1$, respectively. The system size is $N=16$ and $N=8$ in (a) and (b), respectively.
The inset of (a) shows the overlaps of $\rho_{A}$ with different charge sectors ranging from $\{0, \ldots, N/4\}$.  
The QME is observed in U(1)-symmetric random circuits while it is absent in random circuits without any symmetry. Figure reprinted from Ref.~\cite{liu2024symmetry}.}
\label{fig:state_asy_circuit_sym}
\end{center}
\end{figure}

As shown in Fig.~\ref{fig:state_asy_circuit_sym}(a), the early-time QME emerges in the U(1)-symmetric random circuit for tilted ferromagnetic states: states with stronger initial asymmetry (larger EA) decay faster than those with weaker asymmetry. This behavior originates from the distinct thermalization rates across different charge sectors. Specifically, states with greater overlap in the $Q_{A}=0$ sector thermalize more slowly due to the restricted Hilbert space dimension (dim = 1). For instance, the $\theta=0.2\pi$ state decays slowest because it dominantly populates $Q_{A}=0$. The QME similarly appears for tilted domain wall states, following the same mechanism. However, tilted antiferromagnetic states display no QME signature, as states with increasing tilt angle $\theta$ have larger overlap with smaller charge sectors. In contrast, Fig.~\ref{fig:state_asy_circuit_sym}(b) shows that the QME vanishes when the circuit comprises only random Haar gates, as no well-defined charge sectors exist in this case. We further investigate the results of other symmetry groups \cite{liu2024symmetry}. The absence of QME persists in $Z_{2}$-symmetric circuits due to the fact that the Hilbert space dimensions associated with the two parity sectors $Q_{A}=1$ and $Q_{A}=-1$ are exactly equal. However, we observe that QME re-emerges in $SU(2)$-symmetric circuits due to the varying sizes of symmetric sectors.

\subsection{U(1)-symmetric (asymmetric) initial states with U(1) non-symmetric random circuit}
We now investigate the early-time dynamics of entanglement asymmetry in U(1) non-symmetric random circuit, where symmetry-breaking effects are introduced via random Haar gates. The unitary evolution operator $U$ (defined in Eq.~\eqref{unitary_onestep}) is thus modified by substituting a fraction of U(1)-symmetric gates with random Haar gates, while the state $|\psi(t)\rangle$ still updates according to Eq.~\eqref{wf_update}. 

\begin{figure}
\begin{center}
\includegraphics[width=1.03\linewidth]{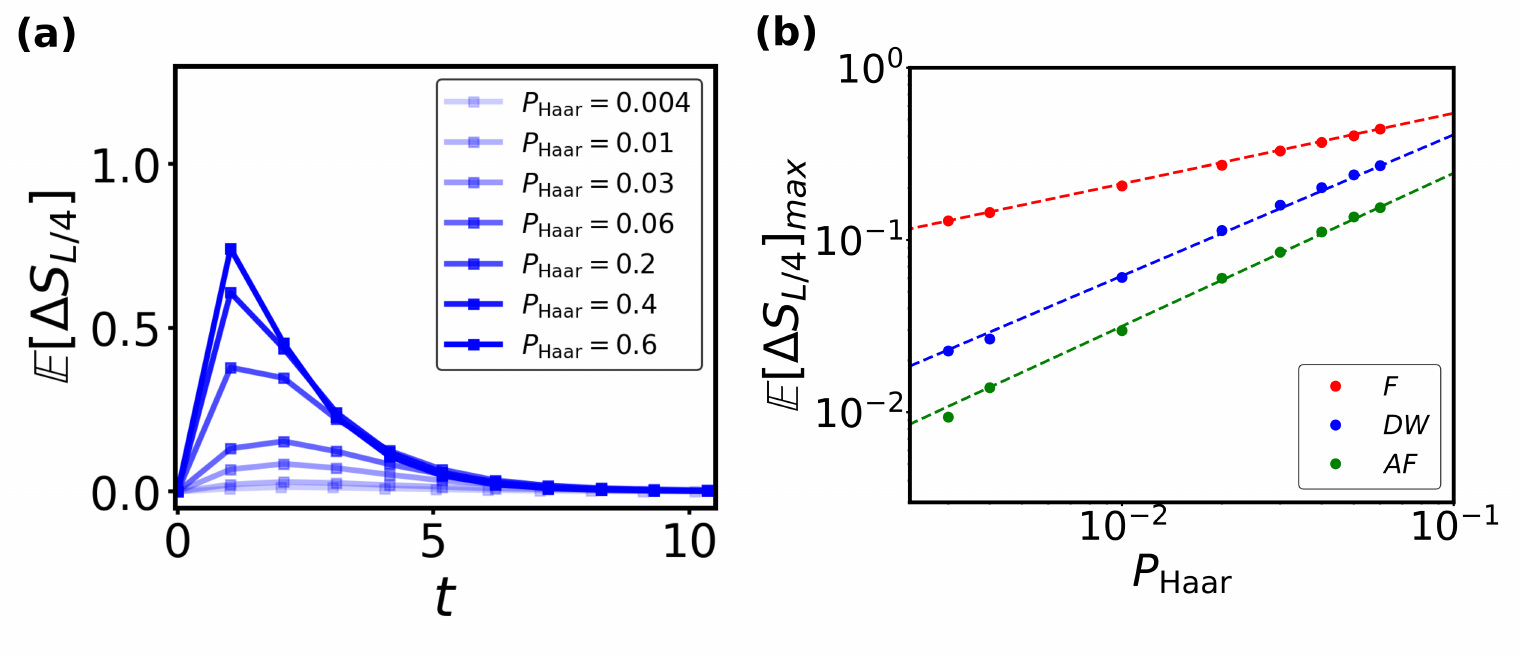}
\caption{(a) The circuit-averaged EA, $\mathbb{E} [\Delta S_{L/4}]$, as a function of time with the antiferromagnetic initial state at different values of $P_{\text{Haar}}$. (b) The peak value, $\mathbb{E} [\Delta S_{L/4}]_{max}$, as a function of $P_{\text{Haar}}$. All three curves follow a power law $y=ax^{b}$. F: Ferromagnetic state ($a=1.4$, $b=0.4$); DW: Domain Wall state ($a=2.7$, $b=0.8$); AF: Antiferromagnetic state ($a=1.9$, $b=0.9$).}
\label{fig:circuit_one}
\end{center}
\end{figure}

We begin by studying the early-time EA dynamics of a U(1)-symmetric initial state evolving under a U(1) non-symmetric random circuit. The circuit under investigation consists of 16 qubits. We evaluate the EA at different doping probabilities of random Haar gates, using an antiferromagnetic initial state. Despite the absence of QME in early-time dynamics, all EAs show nontrivial overshooting at some early time steps, characterized by a peak in EA that significantly exceeds its late-time saturation value. For all probabilities chosen in Fig.~\ref{fig:circuit_one}(a), all EAs reach their maximum after only a few layers of unitaries. The rate of symmetry restoration also depends on the initial state: for antiferromagnetic or domain wall states, symmetry is restored faster than for ferromagnetic states. This difference arises due to the larger Hilbert space sector of the initial states in the former cases. Furthermore, Fig.~\ref{fig:circuit_one}(b) reveals that the peak of the circuit-averaged EA, $\mathbb{E} [\Delta S_{L/4}]_{max}$ follows a power-law with respect to $P_{\text{Haar}}$ across all initial states investigated. This scaling behavior, however, appears to hold only for small $P_{\text{Haar}}<0.1$. 

\begin{figure}[htbp]
\includegraphics[scale=0.53]{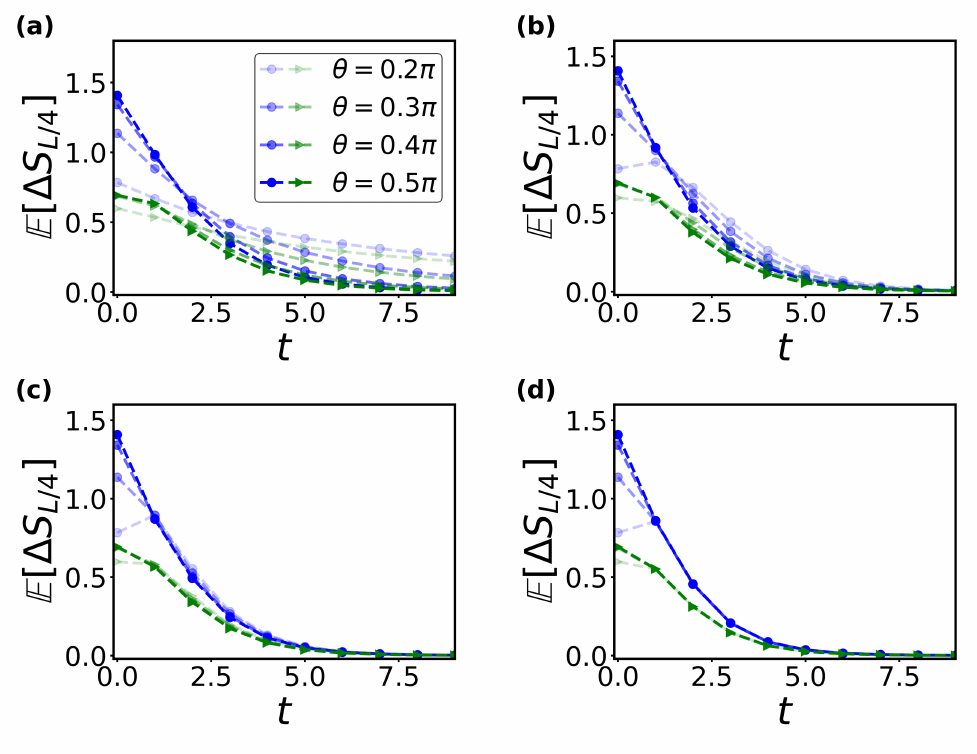}
\caption{The circuit-averaged EA, $\mathbb{E} [\Delta S_{L/4}]$, as a function of time for different values of $P_{\text{Haar}}$. Blue: U(1) EA. Green:  $Z_{2}$ EA. A larger $\theta$ corresponds to a stronger initial symmetry breaking in the system's state. Panels (a)-(d) correspond to different values of $P_{\text{Haar}}$ (a) $P_{\text{Haar}}=0$, (b) $P_{\text{Haar}}=0.3$, (c) $P_{\text{Haar}}=0.7$, and (d) $P_{\text{Haar}}=1$, respectively.}
\label{fig:qme-nonsymm cricuit}
\end{figure}

Next, we examine the evolution of EA for initial states tilted by an angle $\theta$. By systematically varying the parameters such as $P_{\text{Haar}}$ and the initial state $\vert \psi_{\theta} (0) \rangle$, we explore the conditions under which the QME emerges. As shown in Fig.~\ref{fig:qme-nonsymm cricuit}, we compute the EA for both U(1) symmetry with $\hat{Q}_{A} = \sum_{i \in A} \sigma_{i}^{z}$ and $Z_{2}$ symmetry with $\hat{Q}_{A} = \prod_{i \in A} \sigma_{i}^{z}$ for U(1)-asymmetric initial states, i.e. a tilted ferromagnetic state. As depicted in Fig.~\ref{fig:qme-nonsymm cricuit}(a), for $P_{\text{Haar}}=0$, we clearly notice the emergence of QME in the U(1) case. Surprisingly, we also find that the QME appears in the $Z_{2}$ probe, which does not contradict the previous study \cite{liu2024symmetry} suggesting the absence of QME in $Z_{2}$-symmetric circuits. Even though U(1)-symmetric gates are also $Z_{2}$ symmetric, there is no off-diagonal coupling between sectors $\vert 00 \rangle$ and $\vert 11 \rangle$, leading to different thermalization rates between two $Z_{2}$ charge sectors ($Q_{A}=\pm 1$), and thus resulting in QME. 

\begin{figure*}[htbp]
\begin{center}
\includegraphics[scale=0.65]{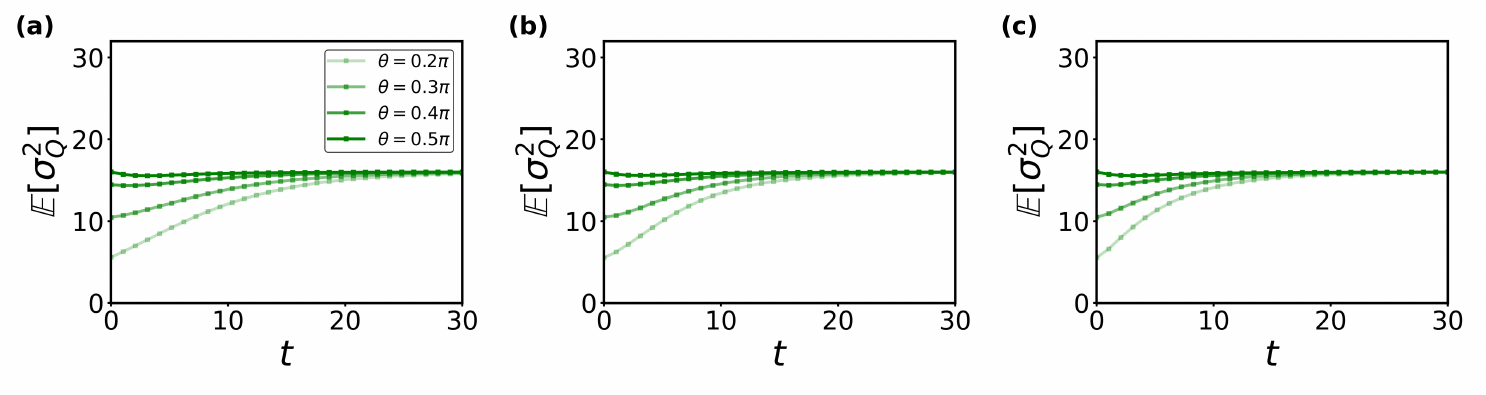}
\caption{Dynamics of circuit averaged CV, $\mathbb{E} [\sigma^{2}_{Q}]$, for different U(1)-asymmetric initial states in U(1) non-symmetric random circuit for $P_{\text{Haar}}=0.05$. The left, middle, and right panels correspond to tilted ferromagnetic, tilted domain wall, and tilted antiferromagnetic initial states, respectively.}
\label{fig:CV_circuit}
\end{center}
\end{figure*}

As we replace a portion of U(1)-symmetric gates with random Haar gates, QME remains evident even with a finite number of random Haar gates.
However, when the circuit consists entirely of random Haar gates, all charge sectors thermalize at the same rate after circuit averaging, and QME disappears. Additionally, initial information, such as the dependence on different $\theta$-values, is erased after applying just one layer of Haar gates. Furthermore, we identify one general property in Fig.~\ref{fig:qme-nonsymm cricuit}. For the same $\theta$ and $P_{\text{Haar}}$, the $Z_{2}$ EA is consistently smaller than U(1) EA. This is because the $Z_{2}$ charge sectors consist of only two sub-blocks, whereas the $U(1)$ sectors involve smaller blocks.

Additionally, we study the early-time EA dynamics for both a tilted domain wall and a tilted antiferromagnetic state. Our findings reveal that QME is present in the tilted domain wall state when $P_{\text{Haar}}$ is less than $1$. For all three initial states under consideration, the EA curves corresponding to different values of the tilt angle $\theta$ converge after the application of a single layer of random Haar gates when $P_{\text{Haar}}=1$, indicating the disappearance of QME.

In parallel with the EA analysis, we also compute the early-time dynamics of charge variance. As illustrated in Fig.~\ref{fig:CV_circuit}, CV for initial states exhibiting a stronger symmetry-breaking effect (large $\theta$) stabilize much more rapidly than for states with a weaker symmetry-breaking effect, consistently across all initial conditions. Consequently, the early-time dynamics of CV reveal no indication of QME in this context. In all the scenarios considered above, we observe that late-time EAs approach zero, irrespective of whether the random circuit exhibits U(1) symmetry or not. This behavior can be understood through the lens of quantum thermalization and information scrambling \cite{chen2024subsystem,hayden2007black,sekino2008fast,lashkari2013towards}. Specifically, for random circuit dynamics, the reduced density matrix of the subsystem converges to a fully mixed state, as long as the subsystem size does not exceed half of the total system size. Similarly, the late-time CV for different initial states also approaches the same saturating value.

In Table~\ref{table:circuit}, we provide a summary of the behavior of entanglement asymmetry and charge variance at early and late times for U(1)-asymmetric states in the circuit model. The quantum Mpemba effect occurs only in the early-time dynamics of entanglement asymmetry for tilted ferromagnetic and tilted domain wall states. Moreover, the entanglement asymmetry and charge variance approach distinct values, which are universal across all initial states, at late times.

\begin{table}[ht]
\centering
\resizebox{0.49\textwidth}{!}{ 
\begin{tabular}{cccc} 
\toprule
& \textbf{Ferromagnetic} & \textbf{Domain Wall} & \textbf{Antiferromagnetic} \\
\midrule
$EA\ (early\ time)$ & crossing  & crossing & no crossing \\
\hline
$CV\ (early\ time)$ & no crossing & no crossing & no crossing \\
\hline
$EA\ (late\ time)$ & 0 & 0 & 0 \\
\hline
$CV\ (late\ time)$ & C & C & C \\
\bottomrule
\end{tabular}
}
\caption{The early- and late-time behavior of entanglement asymmetry (EA) and charge variance (CV) under the evolution of random circuits with $0 \leq P_{\text{Haar}} < 1$. Crossing in EA (CV) means when the time-evolution curves of EA (CV) for states with larger $\theta$ intersect with those for smaller $\theta$ at early times. The constant C independent of $\theta$ can be evaluated as $\mathrm{Tr}(\rho Q^{2})-\mathrm{Tr}(\rho Q)^{2}$, where $\rho=\frac{I}{2^{L}}$ is the late-time density matrix for the subsystem and $I$ is an $2^{L}\times 2^{L}$ identity matrix.}
\label{table:circuit}
\end{table}

\subsection{U(1)-asymmetric initial states with U(1) symmetric Hamiltonian}
In Hamiltonian dynamics, the initial state $\vert \psi(0) \rangle$ undergoes unitary evolution, leading to the time-evolved state $e^{-iHt}\vert \psi(0)\rangle$. The Hamiltonian $H$ governing this evolution is defined as:
\begin{eqnarray}
H = & -\frac{1}{4} \sum_{j=1}^{L_{1}} \Big[ \sigma_j^x \sigma_{j+1}^x + \gamma \sigma_j^y \sigma_{j+1}^y + \Delta \sigma_j^z \sigma_{j+1}^z \Big]  \label{eq:Ham} \\
& -\frac{J_{2}}{4} \sum_{j=1}^{L_{2}} \Big[ \sigma_j^x \sigma_{j+2}^x + \sigma_j^y \sigma_{j+2}^y + \Delta_2\sigma_j^z \sigma_{j+2}^z \Big], \notag
\end{eqnarray}
where $\Delta$ and $\Delta_{2}$ are the coefficients for nearest-neighbor and next-nearest-neighbor interactions, respectively. The term $\Delta_{2}$ introduces non-integrability into the system, while $\gamma$ controls the strength of anisotropy, breaking the system's $U(1)$ symmetry when $\gamma\neq1$. Here we focus on the isotropic case ($\gamma=1$), where the U(1) symmetry associated with spin conservation along the z-axis is preserved. Open boundary condition requires $L_{1}=L-1$ and $L_{2}=L-2$. In contrast, periodic boundary conditions enforce $L_{1}=L_{2}=L$ with $\sigma_{L+1}^{\alpha}=\sigma_{1}^{\alpha}$ and $\sigma_{L+2}^{\alpha}=\sigma_{2}^{\alpha}$ for $\alpha=x,y,z$.      

\begin{figure}[htbp]
\includegraphics[scale=0.23]{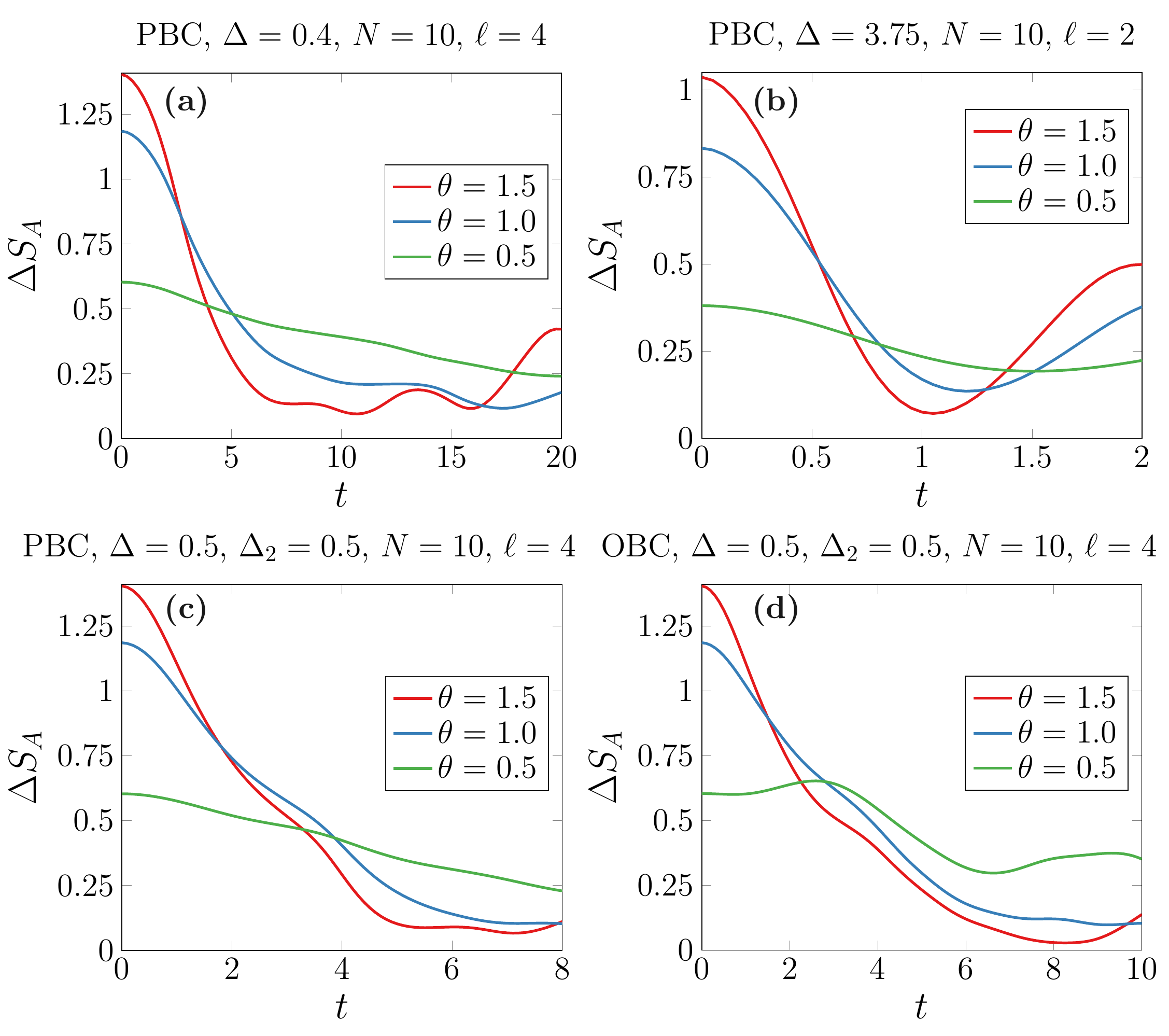}
\caption{The time evolution of the entanglement asymmetry $\Delta S_{A}$ is studied for different tilted ferromagnetic states with system size $L=10$. Panels (a) and (b) correspond to an integrable Hamiltonian with $J_{2}=0$ under periodic boundary conditions (PBC). In contrast, panels (c) and (d) depict results for a non-integrable Hamiltonian with $J_{2}=1$, examining both periodic and open boundary conditions (OBC). Figure reprinted from Ref.~\cite{ares2023entanglement}.}
\label{fig:ham_QME}
\end{figure}

\begin{figure}[htbp]
\begin{center}
\includegraphics[height=2.1cm,width=8.5cm,keepaspectratio=false]{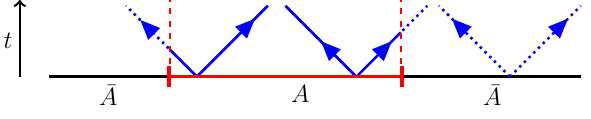}
\caption{Quasi-particle interpretation of entanglement asymmetry. The system is partitioned into subsystem $A$ and its complement $\overline{A}$. Entangled quasi-particle pairs with opposite momentum are emitted throughout the system. Only quasi-particles within $A$ contribute to the entanglement asymmetry (solid blue lines). Those that leave $A$ or reside entirely in $\overline{A}$ (dashed blue lines) make no contribution to the EA. Figure reprinted from Ref.~\cite{ares2025quantum}.}
\label{fig:quasi_QME}
\end{center}
\end{figure}

Fig.~\ref{fig:ham_QME} shows the time evolution of entanglement asymmetry for both integrable and non-integrable Hamiltonians under open and periodic boundary conditions. We observe that initial states with higher asymmetry ($\theta=1.5$) relax faster than those with less asymmetry ($\theta=1$) in all cases, demonstrating the emergence of QME and its broad relevance. However, a fundamental question arises: What is the microscopic origin of QME? An explanation has been provided for one-dimensional integrable quantum systems. Integrable quantum systems possess infinitely many conserved quantities, leading to unique non-equilibrium dynamics governed by quasiparticle excitations \cite{calabrese2005evolution,alba2017entanglement,alba2018entanglement}. These excitations become the essential ingredient to explain the QME. In a quantum quench, the initial state acts as a source of quasiparticle pairs emitted homogeneously, propagating at fixed velocities. As illustrated in Fig.~\ref{fig:quasi_QME}, entangled pairs from the same point spread correlations, while their separation reduces entanglement asymmetry in a subsystem $A$. When one quasiparticle leaves $A$, its contribution to EA vanishes, leading to symmetry restoration. The QME emerges from two key factors: (1) more asymmetric initial states contain more symmetry-breaking quasiparticle pairs, and (2) these pairs propagate at state-dependent velocities. When the dominant symmetry-breaking pairs in a highly asymmetric state happen to be faster, they exit $A$ more quickly, causing faster EA decay. This explains why greater initial asymmetry can lead to faster relaxation. The exact conditions for QME depend only on the initial quasiparticle density and their velocities - quantities computable via Bethe Ansatz for generic integrable systems \cite{rylands2024microscopic}. In terms of generic non-integrable chaotic systems, the mechanism for understanding QME is similar to what we discussed in symmetric circuit cases where relaxation speeds are different for symmetry sectors of different dimensions.

The late-time EA in this case shows two distinct behaviors in the thermodynamic limit, depending on the nature of the quantum system. In chaotic systems, EA vanishes in the thermodynamic limit. This behavior stems from the thermalization properties of chaotic quantum systems: when a closed quantum system evolves under a chaotic Hamiltonian, the reduced density matrix of a small subsystem $A$ thermalizes to the equilibrium finite-temperature state: $\rho_{A} \propto e^{-\beta \hat{H}_{A}}$ where $\hat{H}_{A}$ is the Hamiltonian of the subsystem \cite{deutsch1991quantum,srednicki1994chaos,rigol2008thermalization}. Symmetry is restored at later times for symmetric Hamiltonian $\hat{H}_{A}$, since $[\hat{Q}_{A}, \rho_{A}]=0$ where $\hat{Q}_A$ represents the corresponding symmetry generator. In contrast, the value of steady-state EA is not necessarily zero for integrable systems where the late-time states are expected to be the generalized Gibbs ensemble \cite{rigol2007relaxation,vidmar2016generalized,essler2016quench,caux2013time,ilievski2015complete}, as identified for tilted antiferromagnetic initial states \cite{ares2023lack}.

\subsection{U(1)-asymmetric initial states with U(1)-symmetric many-body localized Hamiltonian}
Symmetry restoration and QME have also been studied in MBL systems \cite{liu2024quantum} which also respects the U(1) symmetry as particle conservation. Unlike integrable or chaotic systems, where QME depends on the initial state, MBL phases exhibit QME universally for any tilted product state. Notably, the timescale for QME grows exponentially with subsystem size, reflecting the logarithmic lightcone nature of MBL dynamics \cite{abanin2019colloquium,pal2010many,nandkishore2015many,imbrie2017local,altman2015universal,lukin2019probing,bardarson2012unbounded,deng2017logarithmic,huang2017out,fan2017out,chen2017out,banuls2017dynamics} . Remarkably, the subsystem symmetry fully restores in the long-time limit, even though the steady state never thermalizes. These findings reveal a unique mechanism in MBL systems, distinct from thermal or integrable regimes, where symmetry restoration and QME are present without thermal equilibrium. The underlying mechanism in MBL systems can be analytically understood via the completely diagonalized effective model for MBL \cite{huse2014phenomenology} and the results are qualitatively the same for both random disorder and quasiperiodic potential induced MBL \cite{iyer2013many,khemani2017two,zhang2018universal}.

\section{Symmetry breaking dynamics}
In this section, we focus on quench dynamics with non-vanishing steady-state EA.

\subsection{U(1)-symmetric initial states with U(1) non-symmetric Hamiltonian}
\begin{figure}
\centering
\includegraphics[width=1.05\linewidth, keepaspectratio]{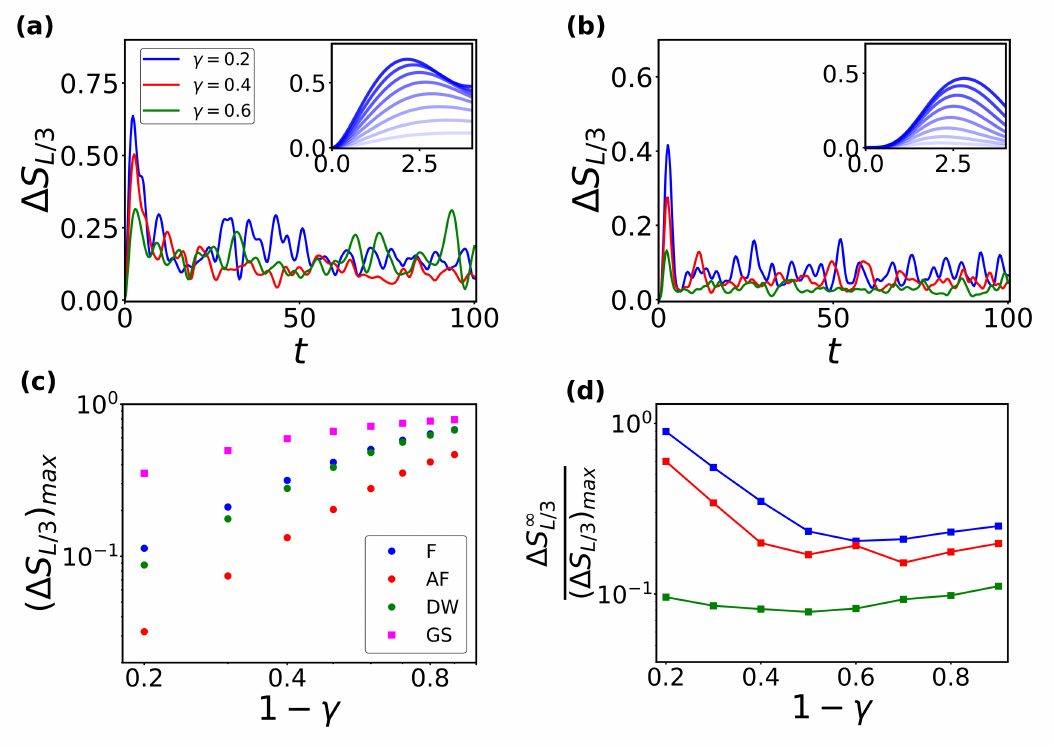}
\caption{EA as a function of time with (a) ferromagnetic and (b) antiferromagnetic states for different values of $\gamma$ under $H_{1}$ with $L=12$. The insets show the peak of EA at different values of $\gamma$. From bottom to top: $\gamma=0.8,0.7,0.6,0.5,0.4,0.3,0.2,0.1$.  Panels (c) and (d) show the peak value of EA, $(\Delta S_{L/3})_{max}$, and the ratio of the late-time EA, $\Delta S_{L/3}^{\infty}$, to $(\Delta S_{L/3})_{max}$ as a function of  $1-\gamma$ for various initial states under $H_{1}$. GS denotes the value of EA calculated from the ground state of $H_{1}$.}
\label{fig:statesym_Hamno}
\end{figure}

We now examine both early- and late-time dynamics of EA and CV in non-symmetric Hamiltonians, investigating whether QME persists in this regime. The symmetry breaking originates from the anisotropy ($\gamma \neq 1$) in Eq.~\eqref{eq:Ham}. We examine two distinct Hamiltonians: (1) $H_{1}$ with parameters $\Delta=0.4$, $J_{2}=0$ (2) $H_{2}$ with parameters $\Delta=0.4$, $J_{2}=0.2$, $\Delta_{2}=1$ for a 12-site system. We begin by investigating the EA dynamics when U(1)-symmetric initial states evolve under these U(1) non-symmetric Hamiltonians. As revealed in Fig.~\ref{fig:statesym_Hamno}(a) and (b), we calculate EA for various Hamiltonian symmetry-breaking values $\gamma$ and observe that EAs also exhibit peaks at early times that are much larger than steady values. Furthermore, the peak value of the EA, $(\Delta S_{L/3})_{max}$, is found to be correlated with the strength of symmetry breaking, $1-\gamma$, for different symmetric initial states as shown in Fig.~\ref{fig:statesym_Hamno}(c) where EA of the ground state of $H_{1}$ follows the same trend. Notably, the peak heights nearly coincide between the ferromagnetic and domain wall states, as the early-time peak primarily depends on the local configurations of the initial state. We conclude that the EA overshooting at early times is a generic feature for asymmetric evolution starting from symmetric initial states.

By analyzing Fig.~\ref{fig:statesym_Hamno}, we identify that the late-time EA, denoted as $\Delta S_{L/3}^{\infty}$, oscillates and does not approach zero. This is because the reduced density matrix of subsystem $A$ evolves towards a thermal equilibrium state $e^{-\beta \hat{H}_{A}}$, where $\hat{H}_{A}$ has the same form as $\hat{H}$ in Eq.~\eqref{eq:Ham}, but acts solely on subsystem $A$. Since $\hat{H}_{A}$ includes symmetry breaking terms, $[\rho_{A},\hat{Q}_{A}]\neq 0$, leading to a non-vanishing EA at long times. In Fig.~\ref{fig:statesym_Hamno}(d), we calculate the ratio of $\Delta S_{L/3}^{\infty}$ to the peak value $(\Delta S_{L/3})_{max}$ with varying $\gamma$ for different initial states. The late-time EA, $\Delta S_{L/3}^{\infty}$, is obtained by averaging $\Delta S_{L/3}$ over $2000$ random time points between $t_{1}=2000$ and $t_{2}=40000$. The results further confirm the overshooting behavior as the late-time saturating EA value is much lower than the peak value at the early time. On the contrary, the CV dynamics in this setting shows no evident overshooting pattern but instead directly grows to the saturating values.

\subsection{U(1) asymmetric initial states with U(1) non-symmetric Hamiltonian}

\begin{figure}[htbp]
\includegraphics[scale=0.48]{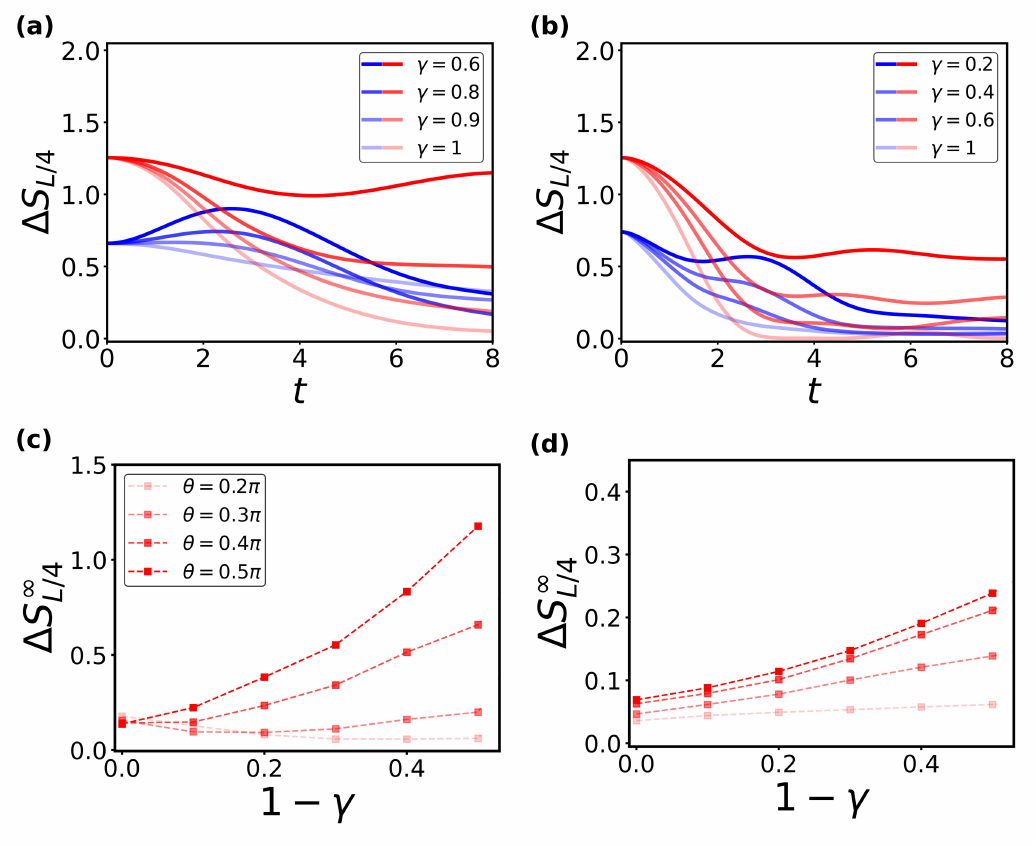}
\caption{Early-time EA dynamics for (a) tilted ferromagnetic states and (b) tilted antiferromagnetic states with varying $\gamma$. The blue curves correspond to $\theta=0.2\pi$, and the red curves represent $\theta=0.5\pi$. Late-time EA, $\Delta S_{L/4}^{\infty}$, as a function of $1-\gamma$ for (c) tilted ferromagnetic states and (d) tilted antiferromagnetic states. All calculations are based on $H_{1}$ with $L=12$.}
\label{fig:ham_QME1}
\end{figure}

We now consider U(1)-asymmetric initial states. As illustrated in Fig.~\ref{fig:ham_QME1}(a) and (b), we observe the emergence of QME at early times in the symmetric Hamiltonian $\gamma=1$. Here, we focus on QME between two initial states, $\theta=0.2\pi$ and $\theta=0.5\pi$. The origin of this QME can be attributed to the relatively small $ZZ$ term and gapless nature of the Hamiltonian \cite{rylands2024dynamical}. As we deviate from the symmetric point ($\gamma=1$), the overall value of EA increases, exceeding the value found in the symmetric case. This behavior aligns with our expectations, as the symmetry-breaking effects now originate from both the initial state and the Hamiltonian governing the dynamics. It is worth noting that the EA curve for initial states with stronger asymmetry rises higher than for states with weaker asymmetry when $1-\gamma\neq 0$, eventually leading to the disappearance of QME. Additionally, the threshold at which QME vanishes varies depending on the type of initial states. QME persists for ferromagnetic states in the range $0.8 \leq \gamma \leq 1$ and for antiferromagnetic states in the range $0.4 \leq \gamma \leq 1$. The robustness of QME against weak symmetry-breaking is a general characteristic of quantum many-body systems when the Hamiltonian breaks the symmetry. 

We also investigate the late-time entanglement asymmetry. As evident from Fig.~\ref{fig:ham_QME1}(c) and (d), the steady-state value of EA exhibits a monotonic increase with the tilt angle $\theta$, consistent with the order at the initial states. This trend persists for tilted antiferromagnetic states even when $\gamma=1$. However, for tilted ferromagnetic initial states at $\gamma=1$, we observe a late-time QME in the EA, where the monotonicity is reversed, i.e. states with smaller $\theta$ now exhibit larger steady state EA.

\begin{figure}[htbp]
\begin{center}
\includegraphics[scale=0.52]{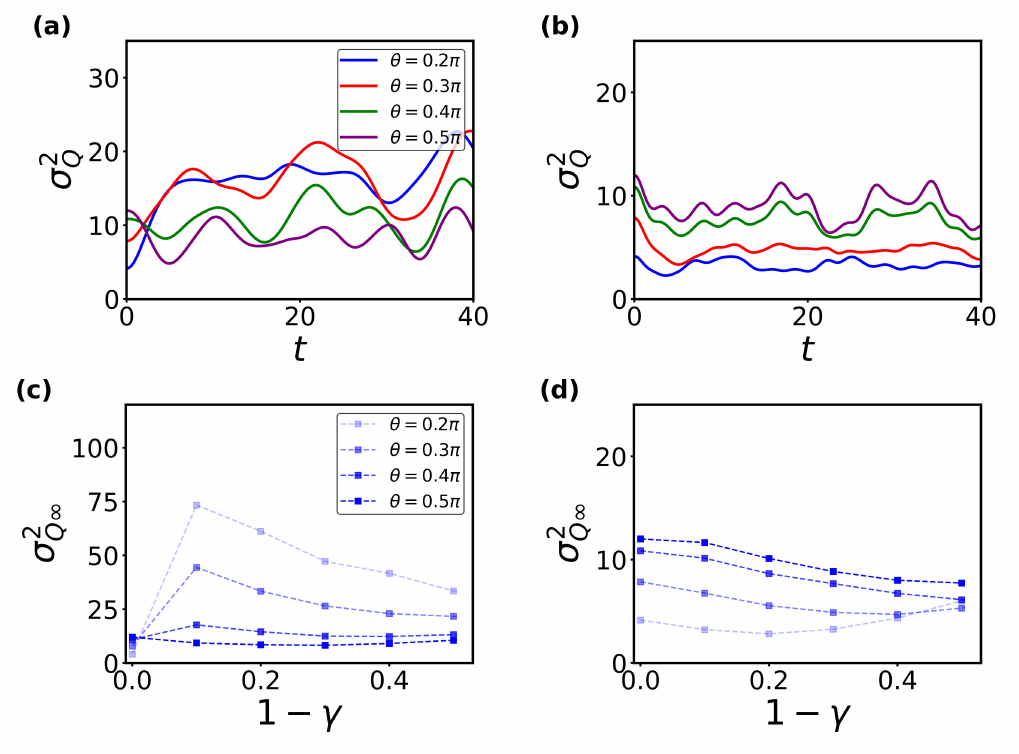}
\caption{Early-time CV dynamics for (a) tilted ferromagnetic states and (b) tilted antiferromagnetic states at $\gamma=0.7$. Curves of different colors correspond to different values of $\theta$. Late-time CV, $\sigma^{2}_{Q \infty}$, as a function of $1-\gamma$ for (c) tilted ferromagnetic states and (d) tilted antiferromagnetic states. All calculations are performed using the Hamiltonian $H_{1}$ with $L=12$.}
\label{fig:Ham_CV}
\end{center}
\end{figure}

Next, we explore the early- and late-time dynamics of charge variance under the evolution of $H_{1}$ and $H_{2}$. As shown in Fig.~\ref{fig:Ham_CV}(b) , for tilted antiferromagnetic states, the charge variance is consistently larger for states with higher asymmetry (large $\theta$) compared to those with lower asymmetry (small $\theta$). However, this trend does not hold for the tilted ferromagnetic initial state, where the monotonic relationship of charge variance and $\theta$ is reversed at early times. As indicated in Fig.~\ref{fig:Ham_CV}(a), this  leads to an early-time crossing around $t=2$, which signifies the presence of QME. To quantitatively analyze the early-time dynamics, we expand the charge variance up to the second order in $t$ as \begin{eqnarray}
\sigma_{Q}^{2}(t) \approx \sigma_{Q}^{2}(0)+\frac{t^{2}}{2}\left. \frac{d^{2}\sigma_{Q}^{2}}{dt^{2}} \right|_{0} ,\label{Cvt expansion}
\end{eqnarray} 
where the linear term in $t$ vanishes. Here, $\left. \frac{d^{2}\sigma_{Q}^{2}}{dt^{2}} \right|_{0}$ represents the second derivative of the charge variance evaluated at $t=0$. The initial state is chosen as a tilted ferromagnetic state. Using the Heisenberg equation of motion, the second derivative of the charge variance at $t=0$ can be expressed as:
\begin{eqnarray}
\left.\frac{d^{2}\sigma_{Q}^{2}}{dt^{2}} \right|_{0} = -\langle [H,[H,Q^{2}]] \rangle_{0} + 2\langle Q \rangle_{0} \langle [H,[H,Q]]\rangle_{0}, \label{all_2nd}
\end{eqnarray} 
where $H$ is the Hamiltonian of the system, taken here as $H_{1}$. The notation $\langle \cdot\rangle_{0}$ denotes the expectation value with respect to the initial state. Without presenting the full derivation, we state that the second derivative of the charge variance at $t=0$ is given by:
\begin{equation}
\begin{split}
\left.\frac{d^{2}\sigma_{Q}^{2}}{dt^{2}} \right|_{0} &= L(1-\gamma)^{2}+3L(1-\gamma)\sin^{2}{\theta}\cos^{2}{\theta} \\ 
&- 3\Delta L(1-\gamma)\sin^{2}{\theta}\cos^{2}{\theta} \\
&+\Delta L(1-\gamma)\sin^{2}{\theta}-L(1-\gamma)\sin^{2}{\theta}.
\end{split}
\end{equation}
The early-time growth of the charge variance for different values of $\theta$ is then approximated by:
\begin{equation}
\begin{split}
\frac{\sigma_{Q}^{2}(t)}{L} &= \sin^{2}{\theta} + \frac{t^{2}}{2}(1-\gamma) \biggl[ (1-\gamma) \\
&\quad + (1-\Delta) \left(3\sin^2 \theta \cos^2\theta - \sin^2\theta \right) \biggr]. \label{complete eq}
\end{split}
\end{equation}
Here, the initial charge variance, $\sigma_{Q}^{2}(0)$, is replaced with $L(1-\cos^{2}{\theta})$. The QME observed in the charge variance for the tilted ferromagnetic state at early times arises from the $t^{2}$ coefficient in Eq.~\eqref{complete eq}, which governs the initial CV dynamics. This term drives a crossing between states with different symmetry-breaking effects. For small tilt angles ($\theta\rightarrow0$), the CV initially grows concavely, characterized by a positive second time derivative, $\frac{d^{2}\sigma_{Q}^{2}}{dt^{2}}>0$. In contrast, as $\theta$ approaches $\pi/2$, the second derivative becomes negative (for suitable values of $\gamma$ and $\Delta$), resulting in a convex decay of the CV at early times. This qualitative difference in dynamical behavior—concave versus convex dynamics—underlies the emergence of the QME in the charge variance. 

\begin{figure}[ht]
\centering
\includegraphics[scale=0.38]{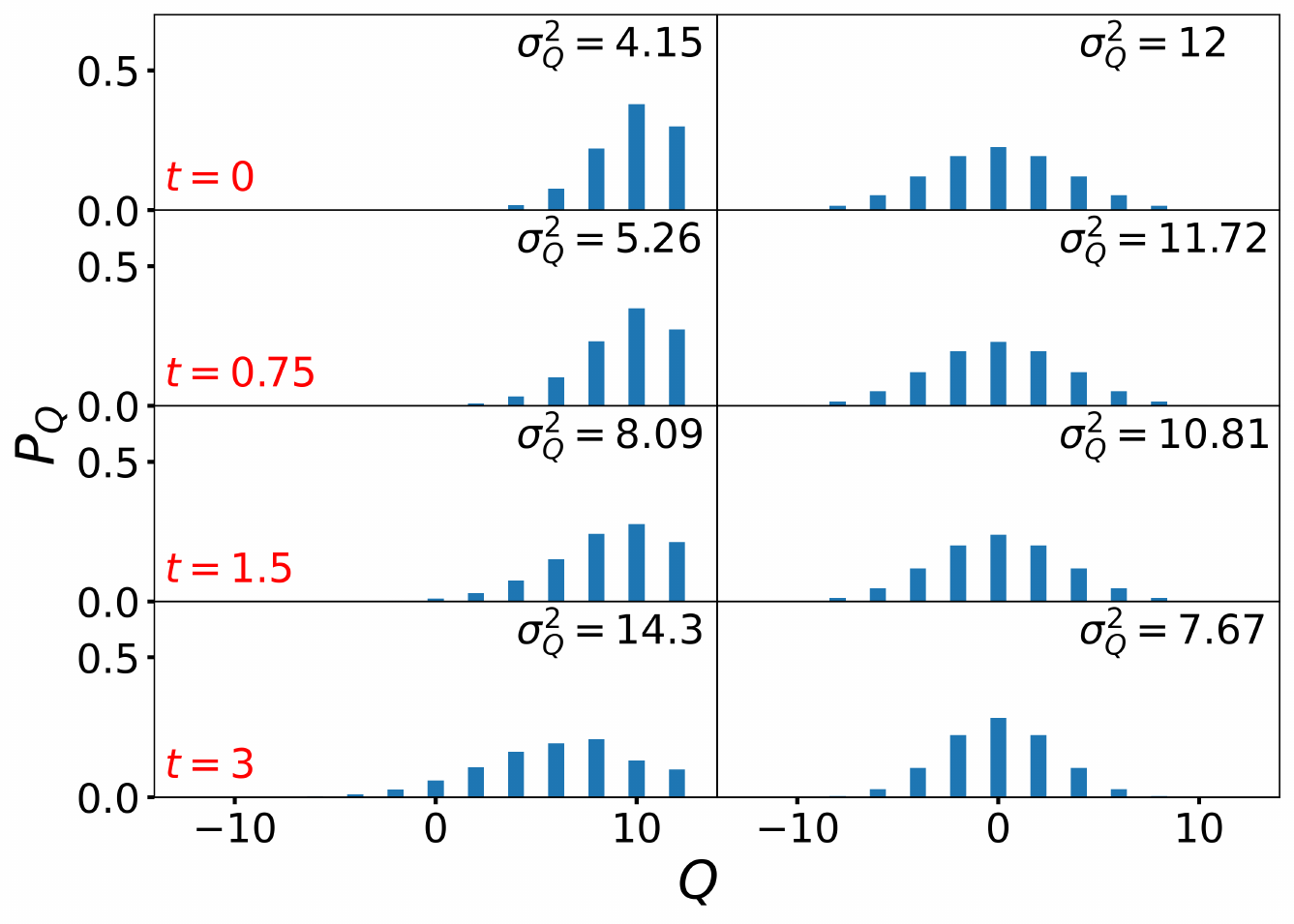}
\caption{Time evolution of the probability distribution, $P_{Q}$, for each charge sector $Q$ under the Hamiltonian $H_{1}$ with $\gamma=0.6$. The charge sectors range from $Q=-12$ to $Q=12$ in increments of $2$. The rows, from top to bottom, correspond to time points $t=0$, $0.75$, $1.5$, and $3$. Two columns, from left to right, represent different tilted ferromagnetic states with $\theta=0.2\pi$ and $0.5\pi$.}
\label{fig:Ham_Chargesector}
\end{figure}
Since the charge variance provides a quantitative measure of the spreading of charge across different sectors $Q$, we analyze the time evolution of the probability distribution $P_Q$ for the state $|\psi(t)\rangle$, initialized in tilted ferromagnetic states. The probability distribution within each sector is defined as:
\begin{eqnarray}
P_{Q}(t) = \sum_{q=Q}|\langle \psi(t)|\psi_{q}\rangle|^{2},
\end{eqnarray}
where $|\psi_{q}\rangle$ denotes the complete set of charge eigenstates with charge $q$. Fig.~\ref{fig:Ham_Chargesector} reveals distinct dynamical behavior for different initial conditions: more asymmetric states (larger $\theta$) exhibit a progressive narrowing of the charge distribution, resulting in suppressed charge variance. Conversely, weakly asymmetric initial states lead to significant broadening of $P_{Q}(t)$ across multiple charge sectors, thereby enhancing the charge variance. This contrasting behavior directly explains the emergence of QME in the early-time charge variance dynamics. To study the late-time behavior, we numerically compute the long-time CV, $\sigma^{2}_{Q\infty}$, by averaging over $2000$ time points between $t_{1}$ and $t_{2}$. As illustrated in Fig.~\ref{fig:Ham_CV}(d), the charge variance for the antiferromagnetic state retains a monotonic dependence on $\theta$ at late times. However, for the tilted ferromagnetic state under non-symmetric evolution, the early-time QME in the CV persists into the late-time regime, indicating an odd number of crossings for CV curves of different $\theta$ during the dynamics.

\begin{table}[ht]
\centering
\resizebox{0.49\textwidth}{!}{ 
\begin{tabular}{cccc} 
\toprule
& \textbf{Ferromagnetic} & \textbf{Domain Wall} & \textbf{Antiferromagnetic} \\
\midrule
$EA\ (early\ time)$ & \parbox{3cm}{crossing for small $1-\gamma$} & \parbox{3cm}{crossing for small $1-\gamma$} &  \parbox{3cm}{crossing for small $1-\gamma$} \\
\hline
$CV\ (early\ time)$ & \parbox{3cm}{crossing for  $\gamma \neq 1$} & no crossing & no crossing \\
\hline
$EA\ (late\ time)$ & \tikz[baseline]{\draw[->, line width=0.35mm] (-0.1,-0.1) -- (0.2,0.2);} & \tikz[baseline]{\draw[->, line width=0.35mm] (-0.1,-0.1) -- (0.2,0.2);} & \tikz[baseline]{\draw[->, line width=0.35mm] (-0.1,-0.1) -- (0.2,0.2);} \\
\hline
$CV\ (late\ time)$ & \tikz[baseline]{\draw[->, line width=0.35mm] (-0.2,0.2) -- (0.1,-0.1);}  & \tikz[baseline]{\draw[->, line width=0.35mm] (-0.1,-0.1) -- (0.2,0.2);} & \tikz[baseline]{\draw[->, line width=0.35mm] (-0.1,-0.1) -- (0.2,0.2);} \\
\bottomrule
\end{tabular}
}
\caption{The early- and late-time behavior of EA and CV under the evolution of $H_{1}$ or $H_{2}$ ($0.5 \leq \gamma \leq 1$). Crossing in EA (CV) means when the time-evolution curves of EA (CV) for states with larger $\theta$ intersect with those for smaller $\theta$ at early times. The right-up (right-down) arrow indicates that the late-time value is increasing (decreasing) with increasing tilted angle $\theta$.}
\label{Table:EA and CV}
\end{table}
In Table~\ref{Table:EA and CV}, we summarize the early- and late-time behavior of EA and CV for different initial states under evolution governed by the non-symmetric Hamiltonian. Moreover, the results obtained from the non-integrable Hamiltonian $H_{2}$ remain qualitatively consistent with those of $H_{1}$. The contrasting behaviors of EA and CV highlight their complementary roles in characterizing the strength and patterns of symmetry breaking.

\section{Open questions and future directions}
In this review, we systematically investigate the dynamics of subsystem symmetry breaking and restoration, examining the evolution of both symmetric and asymmetric initial states under symmetric and symmetry-broken Hamiltonians or random circuits. Our results demonstrate that persistent symmetry breaking within the subsystem occurs exclusively when the dynamics are driven by a non-symmetric Hamiltonian. This enduring asymmetry manifests in the long-time behavior of the system, as diagnosed by a non-vanishing entanglement asymmetry. We further uncover distinct manifestations of the quantum Mpemba effect across different measures: (1) The entanglement asymmetry version arises under symmetric evolution and remains robust against weak symmetry-breaking perturbations, while (2) the charge variance version emerges only in non-symmetric Hamiltonian dynamics. Despite these observations, several fundamental questions remain rarely explored, presenting key opportunities for future research.

One of the fundamental challenges lies in determining whether discoveries about the quantum Mpemba effect could provide a deeper understanding of its classical version. Investigating how classical and quantum relaxation dynamics share similarities might reveal a unified description of anomalous thermalization, connecting these seemingly distinct phenomena. 

Our previous study \cite{yu2025symmetry} introduces two symmetry-breaking protocols for Hamiltonian and random circuit dynamics. This naturally leads to an important open question: how universal is the quantum Mpemba effect when subjected to diverse symmetry-breaking perturbations? For example, do non-Abelian symmetries, discrete symmetries, and alternative symmetry-breaking mechanisms (boundary impurity \cite{gao2025boundary}, time-dependent perturbations) exhibit anomalous relaxation dynamics? A comprehensive investigation of the effect's robustness across different symmetry-breaking scenarios could yield new insights into non-equilibrium dynamics. 

Beyond its standalone significance, the QME may have deep connections to other anomalous dynamics in quantum many-body systems. For example, how does the QME relate to other anomalous dynamical phenomena, such as dynamical phase transitions \cite{heyl2013dynamical,heyl2018dynamical}, prethermalization \cite{berges2004prethermalization,gring2012relaxation}, the Kibble-Zurek mechanism \cite{zurek1985cosmological,hendry1994generation,kibble1976topology,yin2014nonequilibrium}, discrete time crystal \cite{zhang2017observation,else2020discrete}? Are there common underlying principles or can these phenomena influence each other?

Furthermore, another essential direction for future research is understanding how the QME manifests across a broader range of quantum systems. This includes examining its behavior in systems with long-range interactions, higher spatial dimensions, Floquet and annealing protocols (as opposed to quench dynamics). Such studies could reveal whether the QME is a generic feature of non-equilibrium quantum systems or if its emergence depends critically on specific conditions.

Finally, the quantum Mpemba effect offers practical advantages in fast non-adiabatic state preparation. Unlike slow adiabatic cooling, it provides a shortcut to the targets, even from highly non-equilibrium conditions. This could revolutionize quantum control by enabling faster quantum state preparation in the context of quantum simulation and information processing. 

Despite significant progress in studying the Mpemba effect across various systems, many mysteries remain unresolved. From deep theoretical puzzles to real-world applications, further research could uncover exciting breakthroughs. The ongoing investigation of this phenomenon may lead to surprising insights and innovative technologies in the future.\\

\section*{ACKNOWLEDGMENTS}
We are grateful for the valuable discussions and previous collaborations on related topics with Yu-Qin Chen, Shao-Kai Jian, Zi-Xiang Li, Hong Yao, Shuai Yin, and Hao-Kai Zhang.

\section*{Authors' contributions}
SXZ supervised the project. HY prepared the initial manuscript. All authors contributed to the manuscript revision.

\section*{Funding}
HY is supported by the International Young Scientist Fellowship of Institute of Physics Chinese Academy of Sciences (No.202407). SXZ is supported by the Innovation Program for Quantum Science and Technology and a start-up grant at IOP-CAS.

\section*{Data availability}
Supporting information is available from the authors.

\section*{Competing Interests}
The authors declare that they have no competing interests.

\end{document}